\def\diff{\mathrm{d}}

\def\a{\mathrm{a}}
\def\b{\mathrm{b}}
\def\c{\mathrm{c}}
\def\d{\mathrm{d}}
\def\A{\mathrm{A}}
\def\B{\mathrm{B}}
\def\C{\mathrm{C}}
\def\D{\mathrm{D}}

\def\vecp{\mathbf{p}}
\def\Ntest{N_{\textrm{test}}}

\documentclass[a4paper]{jpconf}
\usepackage{graphicx}
\usepackage{graphicx}\usepackage{epsfig}
\usepackage{amsmath}\usepackage{amssymb}\usepackage{amsfonts}
\begin{document}
\title{Bifurcations in dissipative fermionic dynamics}

\author{Paolo Napolitani$^{1}$, Maria Colonna$^{2}$ and Mariangela Di Prima$^{2}$}

\address{$^{1}$IPN, CNRS/IN2P3, Universit\'e Paris-Sud 11, 91406 Orsay cedex, France\\
$^{2}$INFN-LNS, Laboratori Nazionali del Sud, 95123 Catania, Italy}


\begin{abstract}
The Boltzmann-Langevin One-Body model (BLOB), is a novel one-body transport approach, based on the solution of the Boltzmann-Langevin equation in three dimensions; it is used to handle large-amplitude phase-space fluctuations and has a broad applicability for dissipative fermionic dynamics.
We study the occurrence of bifurcations in the dynamical trajectories describing heavy-ion collisions at Fermi energies.

	The model, applied to dilute systems formed in such collisions, reveals to be closer to the observation than previous attempts to include a Langevin term in Boltzmann theories.
The onset of bifurcations and bimodal behaviour in dynamical trajectories, determines the fragment-formation mechanism.
In particular, in the proximity of a threshold, fluctuations between two energetically favourable mechanisms stand out, so that when evolving from the same entrance channel, a variety of exit channels is accessible.

	This description gives quantitative indications about two threshold situations which characterise heavy-ion collisions at Fermi energies.
First, the fusion-to-multifragmentation threshold in central collisions, where the system either reverts to a compact shape, or splits into several pieces of similar sizes.
Second, the transition from binary mechanisms to neck fragmentation (in general, ternary channels), in peripheral collisions.
\end{abstract}

\section{Introduction}
\begin{figure}[t]
\includegraphics[width=17pc]{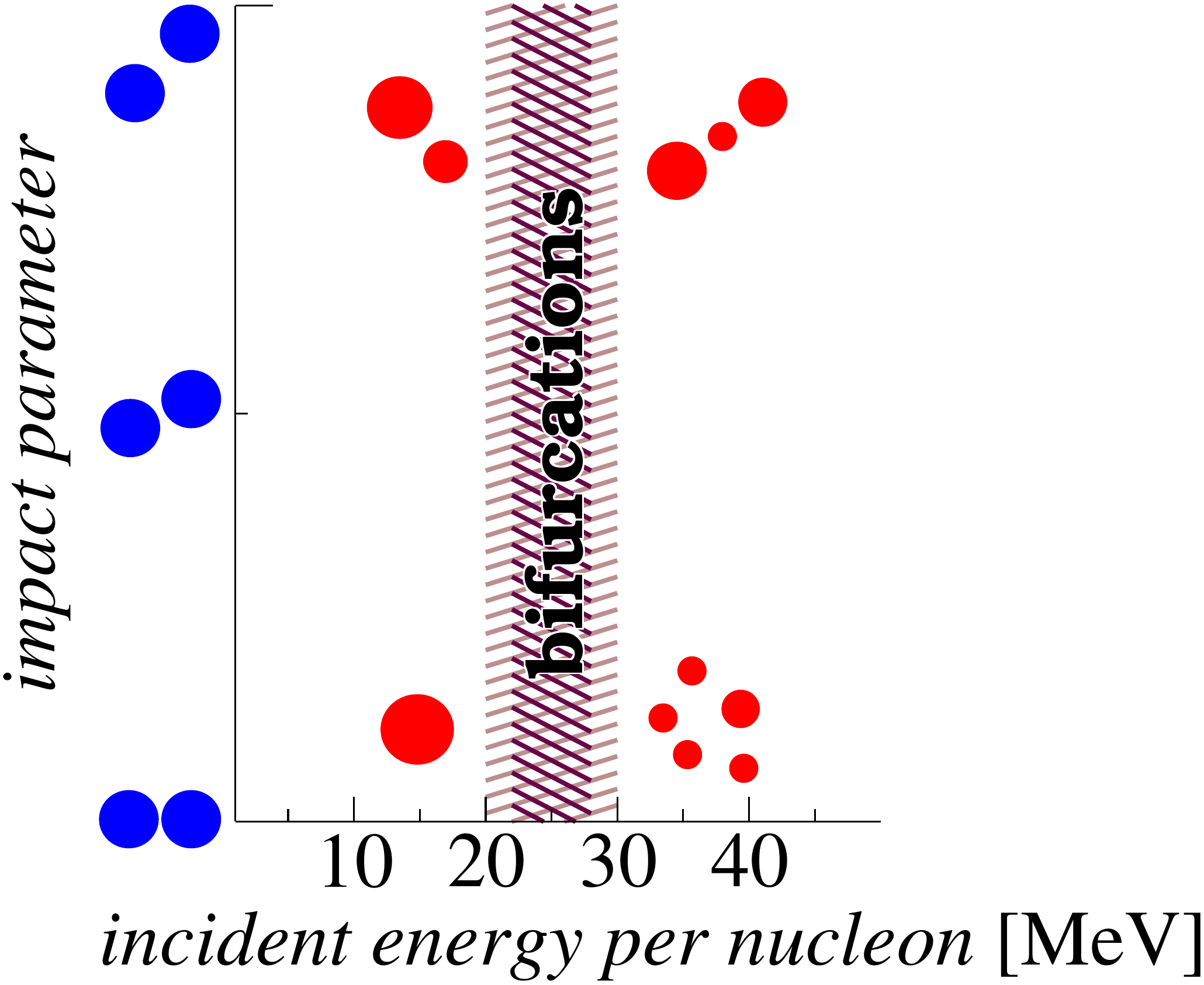}\hspace{2pc}%
\begin{minipage}[b]{19pc}\caption{\label{fig1}Schematic illustration of a variety of exit channels as a function of different entrance channels (impact parameter and incident energy per nucleon): there are regions where the system oscillates between more possible configurations for one given entrance channel.}
\end{minipage}
\end{figure}

\vspace{.5ex}
	The general context of this study is describing the dynamics of fermionic systems in presence of instabilities which produce fluctuations of so large amplitude to produce bifurcations in the dynamical evolution.
	In the case of dissipative heavy-ion collisions, one given projectile-target system evolving from a given entrance channel, defined by an impact parameter and an incident energy, may produce a variety of exit channels.
	For instance, when approaching Fermi energies, heavy-ion collisions may oscillate between fusion and multifragmentation for small impact parameters, or between binary and ternary mechanisms for large impact parameters (fig.~\ref{fig1}).
	To describe such a chaotic behaviour, dominated by fluctuations and bifurcations, pure mean-field equations should be replaced by more adapted theories, like stochastic approaches~\cite{Reinhard1992}.

\begin{figure}[hb]
\begin{minipage}{38pc}
\includegraphics[width=30pc]{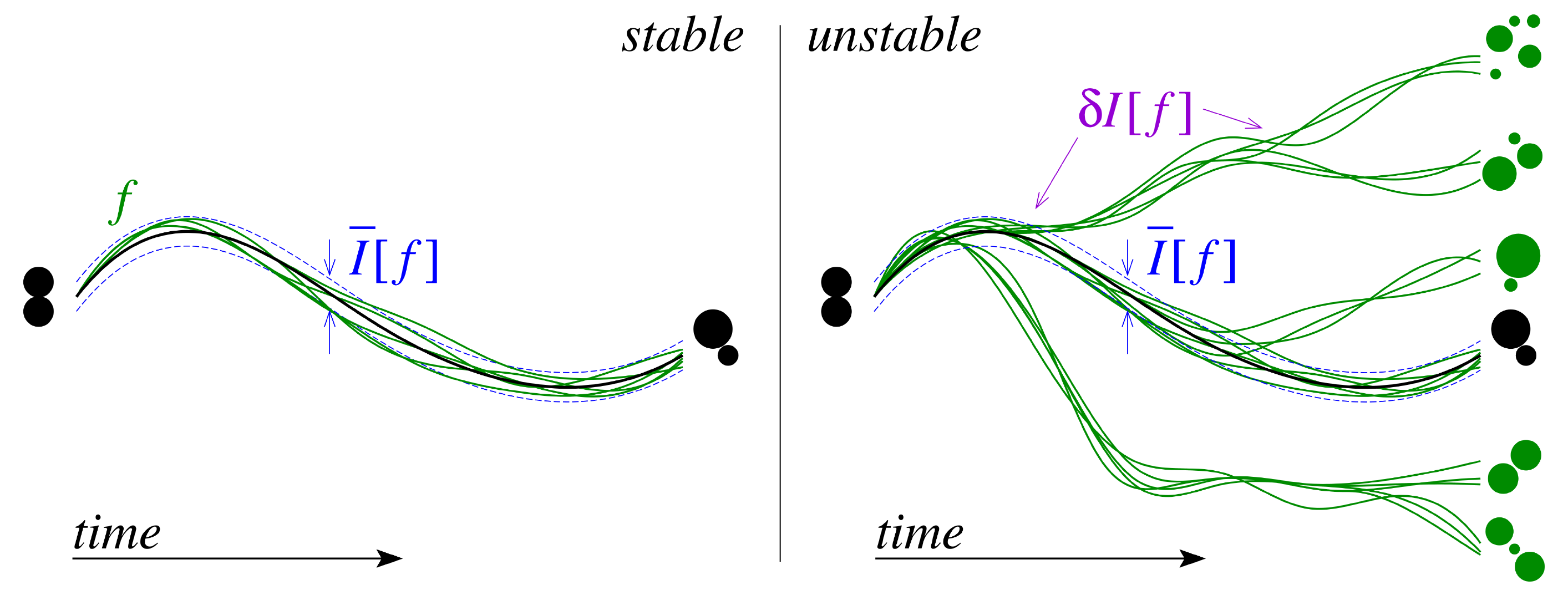}
\caption{\label{fig2}Schematic illustration of the temporal evolution of the one-body distribution function in absence (left) and in presence (right) of fluctuation.}
\end{minipage}\hspace{2pc}
\end{figure}

	Fluctuations are the effect of the correlations which a many-body system generates.
To take them into account there exist two main lanes.
	Either molecular-dynamics approaches, developed from a two-body Hamiltonian.
	Or a hierarchy of N-body contributions, which can be reduced to a mean-field description supplemented by a residual interaction.
	In order to keep an efficient description of effects like spinodal instabilities (relevant in central collisions) and isospin transport~\cite{Baran2005} (like migration toward a neck region, relevant in peripheral collisions), we follow this second approach.
	In this framework, and in a general quantum description, large-amplitude fluctuations in dynamical trajectories may be obtained by defining subensembles of Slater states, each one presenting small fluctuations about its corresponding mean-field~\cite{Balian1986}.
	We adopt a semiclassical analogue of this picture, the Boltzmann-Langevin (BL) transport equation, where the residual interaction carries the unknown N-body correlations and is written in terms of the one-body distribution function as
\begin{equation}
	\partial_t\,f - \left\{H[f],f\right\} = {\bar{I}[f]}+{\delta I[f]} \;,
\label{eq1}
\end{equation}
so that the left-hand side gives the Vlasov evolution for 
the distribution function $f$ in its own self-consistent mean field, and
the right-hand side introduces the residual interaction,
containing the average Boltzmann hard two-body collision integral 
$\bar{I}[f]$ and the fluctuating term $\delta I[f]$.
	This form indicates that the residual interaction, which
carries the unknown N-body correlations is written in terms of the one-body
distribution function.
	The fluctuation term is of Markovian type and it acts as a dissipating force while preserving single-particle energies; it is related to the collision integral through the fluctuation-dissipation theorem in analogy to the Langevin description of the Brownian motion.
	The effect of the fluctuation term on the dynamical trajectories is the possible appearing at any time, whenever the system presents instabilities, of bifurcation branches which propagate in phase space~\cite{Randrup1992} (fig.~\ref{fig2}).

\section{Solution of the Boltzmann-Langevin equation for fermionic systems}
\vspace{.5ex}
	The solution of the BL equation in full phase space is the aim of the Boltzmann-Langevin One-Body model (BLOB)~\cite{Napolitani2013}. It is obtained by replacing the conventional Uehling-Uhlenbeck average collision integral by a similar form where one binary collision does not act on two test particles $a,b$ but it rather involves extended phase-space agglomerates of test particles of equal isospin A$={a_1,a_2,\dots}$, B$={b_1,b_2,\dots}$ (leading to the final states C$={c_1,c_2,\dots}$, D$={d_1,d_2,\dots}$) to simulate wave packets:
\begin{eqnarray}
	{\bar{I}[f]}+{\delta I[f]} 
	&=&\;g\int\frac{\diff\vecp_b}{h^3}\,
	\int \diff\Omega\;\;
	W({\scriptstyle\A\B\leftrightarrow\C\D})\;
	F({\scriptstyle\A\B\rightarrow\C\D}) \\
	&=&\;g\int\frac{\diff\vecp_b}{h^3}\,
	\int \diff\Omega\;\;
	\Big\langle |v_{a}\!-\!v_{b}| \frac{\diff\sigma}{\diff\Omega} \Big\rangle_\Sigma\;
	\Big[(1\!\!-\!\!{f}_\A)(1\!\!-\!\!{f}_\B) f_\C f_\D - f_\A f_\B (1\!\!-\!\!{f}_\C)(1\!\!-\!\!{f}_\D)\Big]
	\,,\notag
\label{eq2}
\end{eqnarray}
	At each interval of time all phase space is scanned for collisions and all test-particle agglomerates are redefined in phase-space cells of volume $h^3$.
	The above procedure introduces correlations which are then exploited through a stochastic procedure, where the effective collision probability $W\!\times\!F$ is confronted with a random number.
	As a consequence, fluctuations develop spontaneously in the phase-space cells  of volume $h^3$ with the correct fluctuation amplitude, determined by a variance which at equilibrium is equal to $f(1-f)$ if the agglomerates contain a number of test particles $\Ntest$ equal to the total number of test particles divided by the number of nucleons constituting the system~\cite{Rizzo2008}.
	Since $\Ntest$ test particles are involved in one collision, and since those test particles could be sorted again in new agglomerates to attempt new collisions in the same interval of time as far as the collision is not successful, the transition rate $W$ in eq.\ref{eq1} should contain a cross section equal to the nucleon-nucleon cross section divided by $\Ntest$: $\sigma = \sigma_{N\!N} / \Ntest$.
	The transition rate 
$W({\scriptstyle\A\B\leftrightarrow\C\D})$
is the average of the elementary transition rates 
$W({\scriptstyle\a\b\leftrightarrow\c\d})$
over the ensemble $\Sigma$ of all the couples of test particles belonging to the agglomerates A and B.

	The BLOB model applies a precise shape-modulation technique~\cite{Napolitani2012} which ensures that the occupancy distribution does not exceed unity in any phase-space point in the final states; this leads to a correct Fermi statistics for the distribution function $f$, in term of mean value and variance.
	The main constraint of the above procedure is to impose a phase-space metric characterised by the phase-space cells  of volume $h^3$, but the metric in momentum and coordinate space are unconstrained in the present approach except for imposing the maximum compactness for the agglomerates in momentum space which does neither violate Pauli blocking nor energy conservation.
	Such strong localisation in momentum space makes the collisions more effective in agitating the phase space.
	However, beyond the present application to Fermi energies, further attention should be paid to the compactness of the wave packets also in coordinate space when dealing with effects like collective flow and stopping, which at intermediate energy become relevant.

\section{Mean-field response in presence of instabilities}
\vspace{.5ex}
If we look for instabilities, we find regions of the equation of state (EoS), like the spinodal region, where the incompressibility $\chi^{-1}=\rho\,\partial P\!/\!\partial\rho$ is negative.
	In such a situation we can explore unstable modes of wave number $k$ growing with a characteristic time $\tau_k$, in relation with the form of the mean-field potential~\cite{Chomaz2004}.

\begin{figure}[h]
\includegraphics[width=18pc]{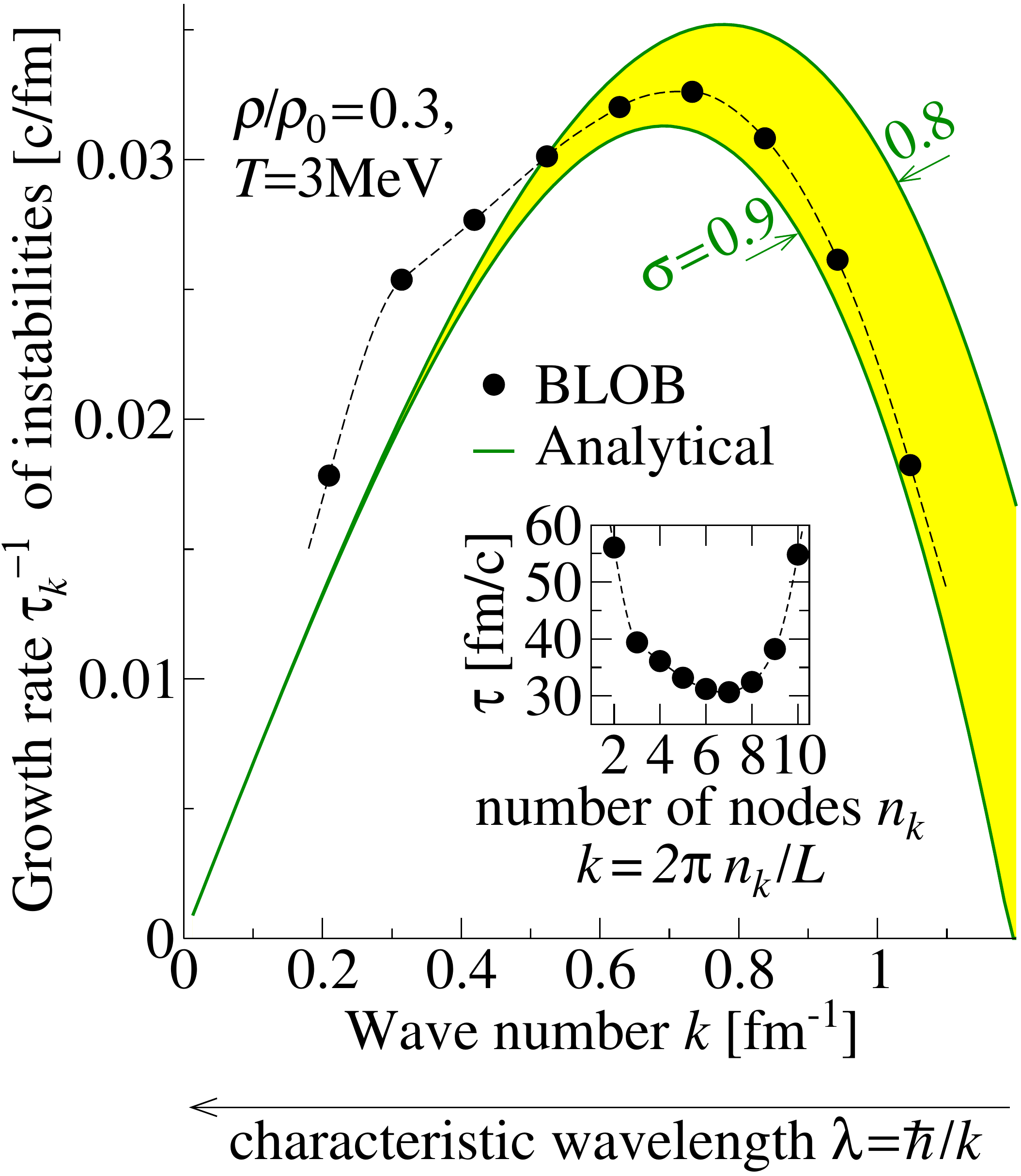}\hspace{2pc}%
\begin{minipage}[b]{18pc}\caption{\label{fig3}Dispersion relation: comparison between an analytical calculation in a linear-response regime (the band) and a corresponding numerical calculation (dots) in a periodic box of size $L$, for a mechanically unstable situation like $T=3$MeV and $\rho/\rho_0=0.05$fm$^{-3}/0.16$fm$^{-3}$. The inset gives the evolution of the characteristic time as a function of the number of nodes.}
\end{minipage}
\end{figure}

	In a linear-response regime we can write, for a three-dimensional system, an analytical expression of the dispersion relation~\cite{Colonna1994} which connects the growth rate $\tau_k^{-1}$ to the form of the mean-field potential as 
$\tau_{k}^{-1}=f(k,\chi,\rho^\prime,T,\sigma)$, where a temperature of $T=3$MeV and a density of $\rho^\prime=0.05$ fm $^{-3}$ (to be compared to the saturation density of 0.16 fm$^{-3}$) are chosen in correspondence with a mechanically unstable situation; the parameter $\chi$ is related to the finite range of the interaction (i.e. the width of the gaussian smearing of the mean-field potential is let range from 0.8 to 0.9 fm, which is consistent with the width of the triangular functions used in the numerical calculation).
	Such analytical calculation is represented by the band in fig.~\ref{fig3}, and it is closely comparable to the result of a transport calculation corresponding to the same mechanically unstable situation.

	The transport calculation is done in a periodic cubic box of edge size $L=60$fm, letting the different modes $k$ develop spontaneously. 
The linear response is calculated tracking the evolution of the density profile along one axis and averaging along the other two transverse coordinates.
The collision term is let agitate the density profile in an initially uniform system
over several wavelengths: from the analysis of this evolution it was deduced how the mean field amplifies each $k$ wave.
The calculation gives the set of points illustrated in the inset of fig.~\ref{fig3}, where the characteristic time is shown as a function of the number of nodes of the wave.
	From these points, the evolution of the growth rate of the instabilities as a function of the wave number is extracted.
	As a consequence of the fact that this numerical approach introduces a coupling among different $k$ values, a shoulder is produced in the spectrum for small $k$ values from the combination of small wavelengths (large $k$) into larger wavelengths (small $k$).
	On the other hand, the analytical calculation does not show any shoulder because, of course, all wavelengths are decoupled.

	Around this efficient description of the dispersion relation, the BLOB model is built and it can therefore be applied to the description of the spinodal behaviour. 
	In this respect, when a nuclear system is brought to small densities (and, correspondingly, large excitation), if spinodal conditions are attained, the system tends to break into fragments with a size $A_{\mathrm{fragm}}\approx \lambda_0^3 \rho_0$, determined by the maximum unstable mode $k_0$ and the corresponding characteristic wavelength $\lambda_0=\hbar/k_0$.

\section{Instability growth versus mean-field resilience}
\vspace{.5ex}
We apply the above phenomenology to the situation of dissipative heavy-ion collisions. 
	For convenience (because experimental data exist, measured by the INDRA collaboration~\cite{Moisan2012}), we simulate the reaction $^{136}$Xe$+^{124}$Sn at 25 and 32$A$MeV with a central impact parameter: in this situation the system is close to a threshold between fusion and multifragmentation (below for 25$A$MeV, above for 32$A$MeV).
	The BLOB transport model is employed with the prescription of a soft EoS 
(with $K_{\mathrm{inf}}=200$MeV), a linear asy-EoS, an in-medium cross section 
as prescribed in ref.~\cite{Danielewicz}, with $\Ntest=40$, and with a statistics 
of 600 events. The dynamics is followed for a time of 300 fm/c.

\begin{figure}[h]
\includegraphics[width=25pc]{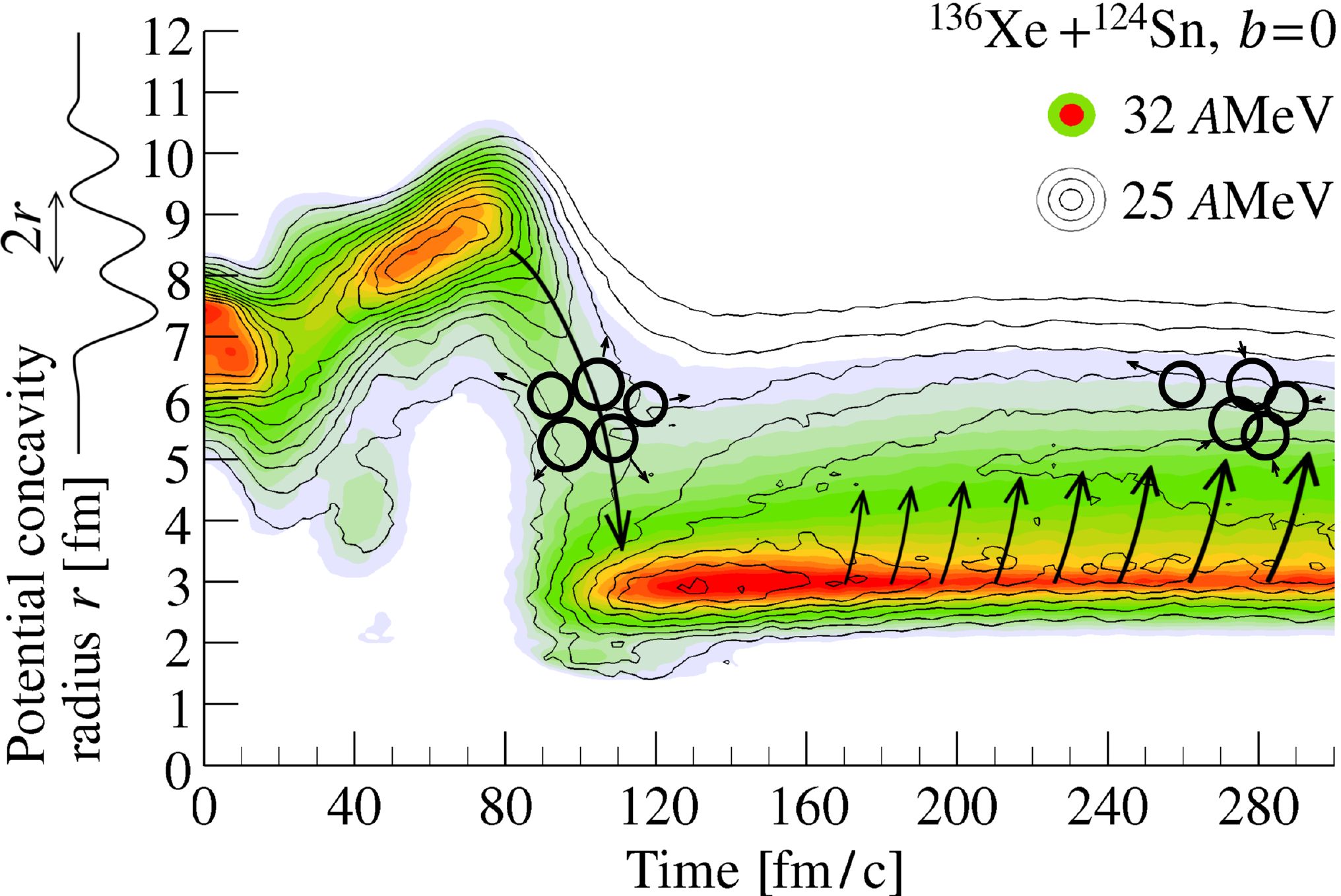}\hspace{2pc}%
\begin{minipage}[b]{11pc}\caption{\label{fig4}Calculation (BLOB) of the evolution of the size of inhomogeneities developing in the potential landscape as a function of time for the reaction $^{136}$Xe$+^{124}$Sn at 32$A$MeV (colours) and 25$A$MeV (contours) for a central impact parameter.}
\end{minipage}
\end{figure}
	Fig.~\ref{fig4} shows the evolution of the size of inhomogeneities which develop in the potential landscape as a function of time for the two incident energies 25 and 32$A$MeV. 
	These inhomogeneities correspond to blobs of matter which, if the dynamics is sufficiently explosive, may eventually leave the system as emitted fragments.
	At short times, as discussed in the previous section, spinodal instabilities tend to split the system into several fragments of comparable size $A_{\mathrm{fragm}}\approx \lambda_0^3 \rho_0$, corresponding approximately to the region of Neon.
	Later on, this process enters in competition with a process of partial coalescence determined by the mean-field resilience, which tends to revert the system to a compact shape, and which is more effective for the lowest incident energy.

	This picture results in a large variety of exit channels for one given macroscopic initial condition (entrance channel), ranging from fusion to multifragmentation and passing through very asymmetric configurations with a small fragment multiplicity, eventually binary, and a large size asymmetry.
	Such intermediate configurations may recall asymmetric fission with the difference that the distribution of momentum transfer could reach large values more compatible with multifragmentation.

\section{Fusion-multifragmentation competition in central collisions}
\vspace{.5ex}
\begin{figure}[t]
\begin{minipage}{38pc}
\includegraphics[width=28pc]{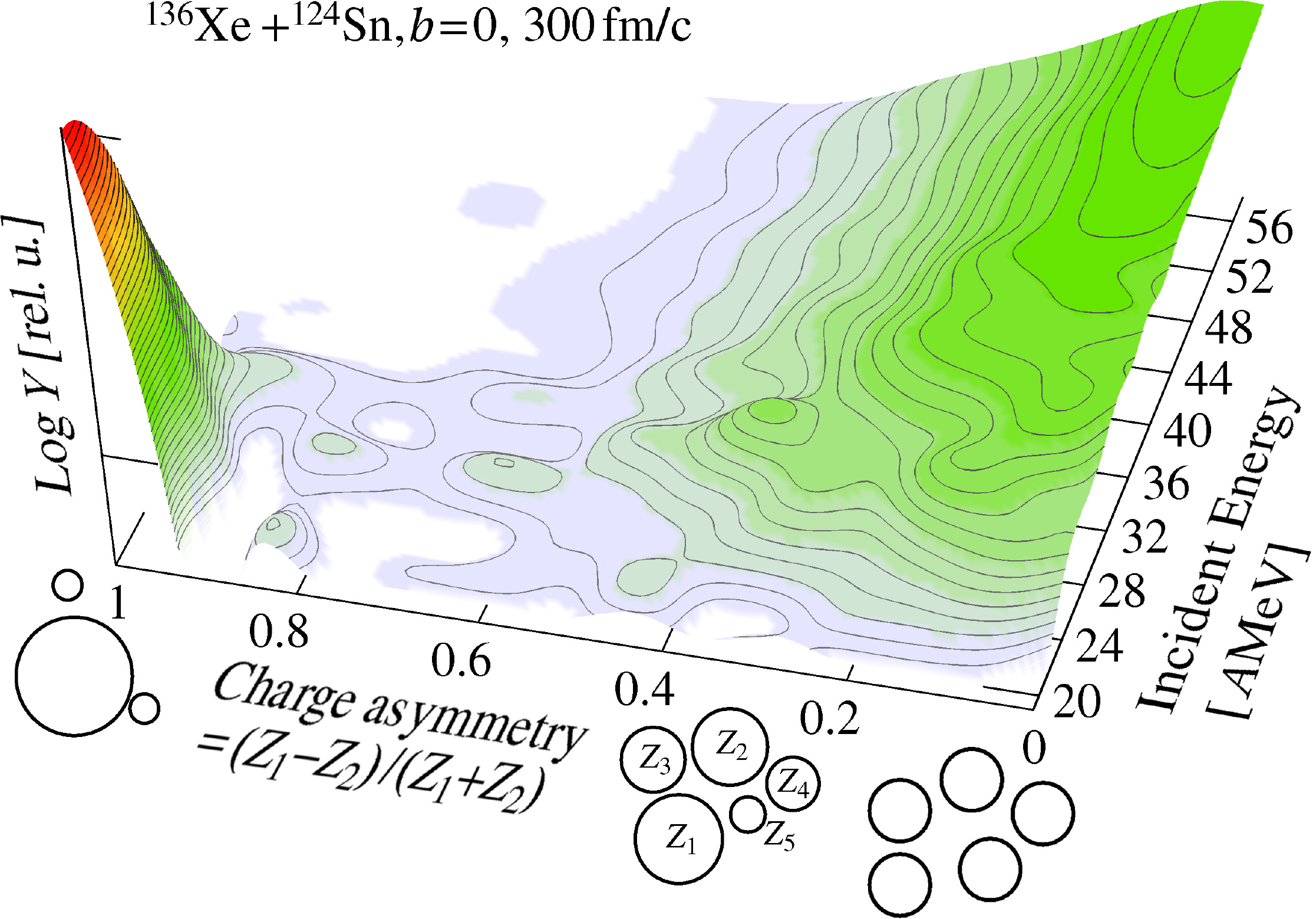}
\caption{\label{fig5}Distribution of exit channels, represented by the charge asymmetry $(Z_1-Z_2)(Z_1+Z_2)$ as a function of the incident energy for the reaction $^{136}$Xe$+^{124}$Sn for a central impact parameter.}
\end{minipage}\hspace{2pc}
\end{figure}
	The distribution of different configurations for the exit channel of a given system can be represented by the charge asymmetry $(Z_1-Z_2)(Z_1+Z_2)$ between the fragment of largest charge $Z_1$ and the fragment of second largest charge $Z_2$.
	If the same calculation of fig.~\ref{fig4} is repeated for a range of different incident energies we obtain the distribution of exit channels illustrated in fig.~\ref{fig5}.
	We observe that for small incident energies the fusion channel is favoured, while for larger incident energies multifragmentation dominates.
	In between, a region exist where dynamical trajectories show bifurcations between these two competing channels.
	
	This dynamical picture, where
energy fluctuations result from the same macroscopic initial conditions and lead to oscillations between two energetically favoured configurations,
provides a link to the thermodynamic picture of a first-order phase transition.
	We also mention that very close experimental observations have found bimodalities in fragment observables in heavy-ion collisions at Fermi energies from the analysis of peripheral impact parameters~\cite{Bonnet2009,Pichon2006}.
	The present theoretical study encourages to extend experimental investigations of bimodal behaviour in fragment observables also to central collisions.

\section{Overview on a landscape of exit channels}
\vspace{.5ex}
	An even more general survey extends the set of calculations also to the full range of impact parameters, and completes the dynamical calculation with a further decay process of the hot fragments through a statistical evaporation model~\cite{Durand1992}.
	We still keep the same reaction $^{136}$Xe$+^{124}$Sn at 32$A$MeV but we additionally vary the impact parameter $b$.
\begin{figure}[ht]
\begin{minipage}{38pc}
\includegraphics[width=37pc]{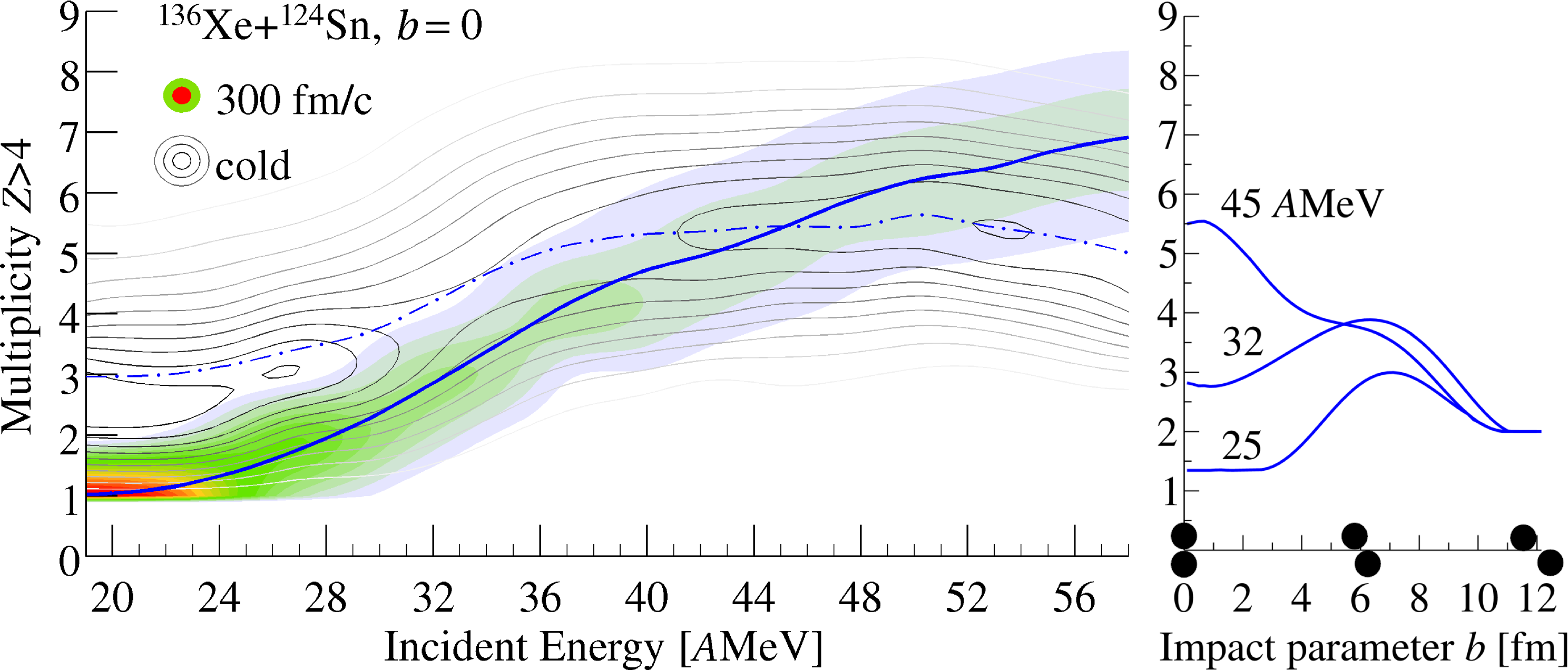}
\caption{\label{fig6}Multiplicity of fragments larger than Be in the reaction $^{136}$Xe$+^{124}$Sn as a function of the incident energy for a central impact parameter (left panel) and as a function of the impact parameter for 25, 32 and 45 $A$MeV (right panel).}
\end{minipage}\hspace{2pc}
\end{figure}

	The left panel of fig.~\ref{fig6} is still restricted to central collisions and shows the evolution of the multiplicity of fragments larger than Be as a function of the incident energy (coloured distribution with the average marked by a solid line). The distribution of the fragments resulting from the evaporation process reaches a maximum above about 35$A$MeV and a fall above 50$A$MeV due to the more and more reduced size of the fragments.
	The right panel of fig.~\ref{fig6} shows the average dependence of the fragment multiplicity as a function of the impact parameter.
	We observe that the maximum corresponds to central collisions for large incident energies because the multifragmentation channel dominates the reaction cross section.
	For smaller incident energies the maximum moves to more peripheral collisions and it is reduced to events with three or four fragments: this indicates that around 30$A$MeV and for semiperipheral impact parameters ($\sim6$fm) the system oscillates between binary mechanisms and ternary splits, where a third smaller fragment is formed in the neck region.

\section{Conclusions}
\vspace{.5ex}
	With the purpose of studying the link between a given entrance channel and a variety of possible exit channels in situations where a nuclear system experiences instabilities, we built a one-body description based on the solution of the Boltzmann-Langevin equation in three dimensions.
	The method is able to introduce fluctuations of correct amplitude in terms of dispersion relation.

	Thanks to this transport model we could draw a dynamical description of the origin of trajectory bifurcations and bimodality in head-on collisions down to the multifragmentation threshold.
	In particular, we observe that this phenomenology results from the competition between instability growth and the resilience of the mean field.
	Within the same picture we might also explain asymmetric binary channels.
Analysing the variation of the exit channel with the impact parameter it is also possible to determine the most favourable conditions for producing fragments from the neck regions in peripheral collisions.

\end{document}